\documentclass[apj]{emulateapj}

\usepackage{amsmath}
\usepackage{txfonts}

\eqsecnum

\renewcommand\email\texttt

\newcommand\coords{{13:28:03.5\, +33:33:21.0}}
\newcommand\galcoords{{74.3,\, 79.8}}
\newcommand\starnum{{J132755.56+333521.7}}

\def\spose#1{\hbox to 0pt{#1\hss}}
\def\lta{\mathrel{\spose{\lower 3pt\hbox{$\sim$}}
    \raise 2.0pt\hbox{$<$}}}
\def\gta{\mathrel{\spose{\lower 3pt\hbox{$\sim$}}
    \raise 2.0pt\hbox{$>$}}}

\begin{document} 

\slugcomment{\sc submitted to \it Astrophysical Journal Letters}
\shorttitle{\sc Canes Venatici Dwarf} 
\shortauthors{Zucker et al.}

\title{A New Milky Way Dwarf Satellite in Canes Venatici}

\author{D.\ B. Zucker\altaffilmark{1}, 
V. Belokurov\altaffilmark{1},
N.\ W. Evans\altaffilmark{1}, 
M.\ I. Wilkinson\altaffilmark{1},
M.\ J. Irwin\altaffilmark{1},
T.\ Sivarani\altaffilmark{2},
S.\ Hodgkin\altaffilmark{1},
D.\ M. Bramich\altaffilmark{1}, 
J.\ M. Irwin\altaffilmark{1},
G. Gilmore\altaffilmark{1}, 
B. Willman\altaffilmark{3},
S. Vidrih\altaffilmark{1},
M. Fellhauer\altaffilmark{1},
P.\ C. Hewett\altaffilmark{1}, 
T.\ C. Beers\altaffilmark{2},
E.\ F.\ Bell\altaffilmark{4},
E.\ K.\ Grebel\altaffilmark{5},
D.\ P.\ Schneider\altaffilmark{6},
H.\ J.\ Newberg\altaffilmark{7},
R.\ F.\ G.\ Wyse\altaffilmark{8},
C.\ M.\ Rockosi\altaffilmark{9},
B. Yanny\altaffilmark{10},
R.\ Lupton\altaffilmark{11},
J.\ A.\ Smith\altaffilmark{12},
J.\ C.\ Barentine\altaffilmark{13},
H.\ Brewington\altaffilmark{13},
J.\ Brinkmann\altaffilmark{13},
M.\ Harvanek\altaffilmark{13},
S.\ J.\ Kleinman\altaffilmark{13},
J.\ Krzesinski\altaffilmark{13,14},
D.\ Long\altaffilmark{13},
A.\ Nitta\altaffilmark{13},
S.\ A.\ Snedden\altaffilmark{13}
}

\altaffiltext{1}{Institute of Astronomy, University of Cambridge,
Madingley Road, Cambridge CB3 0HA, UK;{\tt ~zucker,vasily,nwe@ast.cam.ac.uk}}
\altaffiltext{2}{Department of Physics and Astronomy, CSCE: Center for
the Study of Cosmic Evolution, and JINA: Joint Institute for Nuclear
Astrophysics, Michigan State University, East Lansing, MI 48824}
\altaffiltext{3}{Center for Cosmology and Particle Physics, Department of Physics, New York University, 4 Washington Place, New York, NY 10003}
\altaffiltext{4}{Max Planck Institute for Astronomy, K\"{o}nigstuhl
17, 69117 Heidelberg, Germany}
\altaffiltext{5}{Astronomical Institute of the University of Basel, Department of Physics and Astronomy, Venusstrasse 7,CH-4102 Binningen, Switzerland}
\altaffiltext{6}{Department of Astronomy and Astrophysics, Pennsylvania State University, 525 Davey Laboratory, University Park, PA 16802}
\altaffiltext{7}{Rensselaer Polytechnic Institute, Troy, NY 12180}
\altaffiltext{8}{The Johns Hopkins University, 3701 San Martin Drive,
Baltimore, MD 21218}
\altaffiltext{9}{Lick Observatory, University of California, Santa Cruz, CA 95064}
\altaffiltext{10}{Fermi National Accelerator Laboratory, P.O. Box 500,
Batavia, IL 60510}
\altaffiltext{11}{Princeton University Observatory, Peyton Hall, Princeton, NJ 08544}
\altaffiltext{12}{Los Alamos National Laboratory, ISR-4, MS D448, Los Alamos, NM 87545} 
\altaffiltext{13}{Apache Point Observatory, P.O. Box 59, Sunspot, NM 88349}
\altaffiltext{14}{Mt.\ Suhora Observatory, Cracow Pedagogical University, ul.\ Podchorazych 2, 30-084 Cracow, Poland}

\begin{abstract}
In this Letter, we announce the discovery of a new dwarf satellite of
the Milky Way, located in the constellation Canes Venatici.  It was
found as a stellar overdensity in the North Galactic Cap using Sloan
Digital Sky Survey Data Release 5 (SDSS DR5). The satellite's color-magnitude
diagram shows a well-defined red giant branch, as well as a horizontal
branch. As judged from the tip of the red giant branch, it lies at a
distance of $\sim 220$ kpc. Based on the SDSS data, we estimate an
absolute magnitude of $M_V \sim -7.9$, a central surface brightness of
$\mu_{\rm 0,V} \sim 28$ mag arcsecond$^{-2}$, and a half-light radius
of $\sim 8\farcm5$ ($\sim 550$ pc at the measured distance). The outer
regions of Canes Venatici appear extended and distorted. The discovery
of such a faint galaxy in proximity to the Milky Way strongly suggests
that more such objects remain to be found.
\end{abstract}


\section{Introduction}
There are ten known dwarf spheroidal (dSph) companions of the Milky
Way Galaxy. Together with the two dwarf irregulars (the Large and
Small Magellanic Clouds), this makes up all the known satellite
galaxies of the Milky Way. The dSphs have such low surface brightness
that they have often been found serendipitously. For example, while
Sextans~\citep{Ir90} was found as part of an automated search, the intrinsically far more luminous Sagittarius dSph was first identified kinematically from a radial velocity survey of stars in Galactic Center
fields~\citep{Ib95}.

As such galaxies are resolvable into individual stars because of their
proximity, they are detectable as enhancements in the stellar number
density in large photometric surveys. For example, \citet{Ma04}
analyzed overdensities of M giants in the 2MASS All Sky Survey and
claimed the detection of a new disrupted dSph in Canis Major, although
this remains controversial. \citet{Wi05} systematically surveyed $\sim 5800$ square degrees of the Sloan Digital Sky Survey \citep[SDSS;][]{Yo00} and identified a dSph companion to the Milky Way in the
constellation of Ursa Major. Subsequent spectroscopic
observations \citep{Ke05} showed that Ursa Major was a dark matter
dominated dSph. Meanwhile, \cite{Zu04,Zu06} announced the discovery of
Andromeda IX and X, new dSph satellites of M31 found in SDSS data with
similar methods.

Very recently, \citet{Be06b} mapped out the stars satisfying the color
cut $g-r < 0.4$ in almost all of SDSS Data Release 5 (DR5). They dubbed this plot
of the high-latitude Galactic northern hemisphere the ``Field of Streams'', owing to its wealth of prominent stellar substructure. Visible by eye in the plot is
a heretofore undetected enhancement of the stellar density, located in the
constellation Canes Venatici. Here, we show that this corresponds to a
new dSph companion -- the eleventh -- of the Milky Way Galaxy.

\begin{figure*}[t]
\begin{center}
\epsscale{0.90}
\plotone{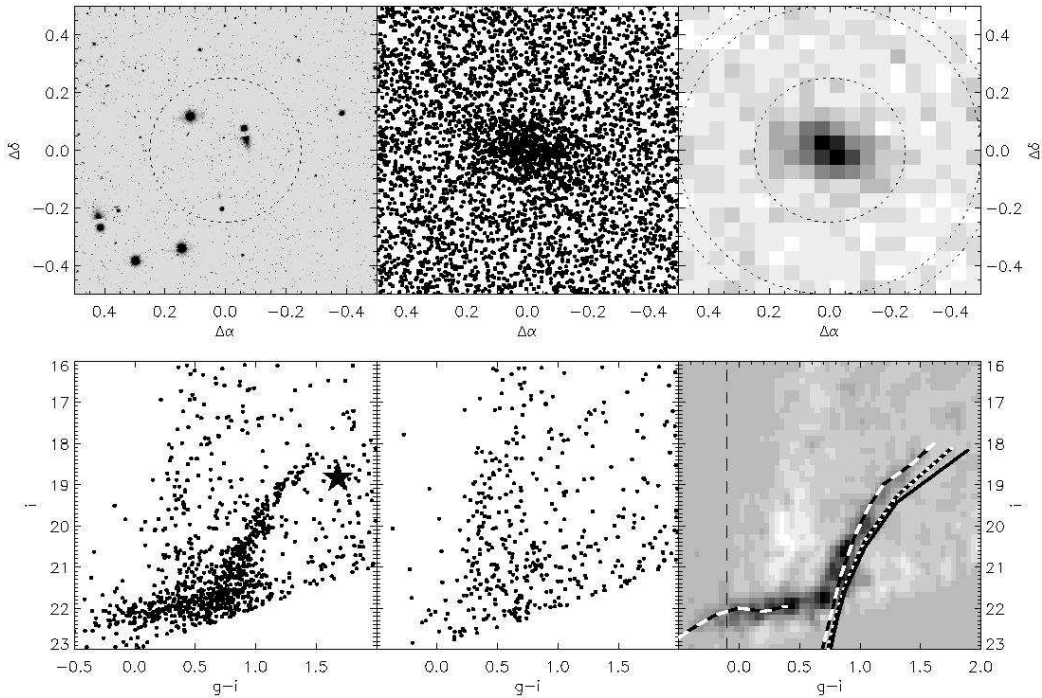}
\caption{The Canes Venatici Dwarf: {\it Upper left:} Combined SDSS
$g,r,i$ images of a $1^{\circ} \times 1^{\circ}$ field centered on the
overdensity. $\Delta \alpha$ and $\Delta\delta$ are the relative offsets in right
ascension and declination, measured in arcdegrees. The dotted circle indicates a radius of $0.25^{\circ}$. {\it Upper middle:}
The spatial distribution of all objects classified as stars in the
same area. {\it Upper right:} Binned spatial density of all stellar
objects. The inner dotted circle marks a radius of $0.25^{\circ}$, the middle circle a radius of $0.5^{\circ}$, and the outermost circle a radius of $0.56^{\circ}$. Bins are $0.05^\circ
\times 0.05^\circ$, smoothed with a Gaussian with a FWHM of
$0.1^\circ$.  {\it Lower left:} CMD of all stellar objects within the
inner $0.25^\circ$ radius circle; note the clear RGB, with hints of a
horizontal branch, even without removal of field contamination. The
filled star shows the location of a carbon star in the field of CVn
(see text). {\it Lower middle:} CMD of a comparably-sized control area, showing all stellar
objects between $0.5^\circ$ (middle circle) and $0.56^{\circ}$ (outer circle) of the
center. {\it Lower right:} A color-magnitude density plot (Hess
diagram), showing the inner CMD minus a control CMD taken from between $0.5^\circ$ and $2^\circ$ of the center, normalized to
the number of stars in each CMD.  RGB and horizontal branch fiducial
ridgelines derived from SDSS photometry of the globular cluster
NGC~2419 ([Fe/H] $\sim -2.1$) are overplotted as dashed
white-and-black lines; RGB fiducials for M~3 ([Fe/H] $\sim -1.6$) and
M~71 ($\sim -0.7$) from \citet{Clem05} are shown in dotted
white-and-black and solid black, respectively. All fiducials were
corrected for Galactic foreground extinction \citep{schl98}, and shifted to a distance modulus of 21.75. The thin dashed vertical line shows the cut $g-i < -0.1$ used
to select candidate blue horizontal branch stars.
\label{fig:cvn_disc}}
\end{center}
\vskip -0.90cm
\end{figure*}

\begin{deluxetable}{lc}
\tablecaption{Properties of the Canes Venatici Dwarf \label{tbl:pars}}
\tablewidth{0pt} \tablehead{ \colhead{Parameter\tablenotemark{a}} &
{~~~ } } \startdata Coordinates (J2000) & \coords $\pm 10\arcsec$\\
 Galactic Coordinates ({\em l,b}) & \galcoords\\
 Position Angle & $73^{\circ} \pm 3^{\circ}$\\
 Ellipticity & $0.38$\\
 $r_h$ (Plummer) & $8\farcm5 \pm 0\farcm5$\\
$r_h$ (Exponential) & $8\farcm4 \pm 0\farcm5$\\
 A$_{\rm V}$ & $0\fm05$ \\
 $\mu_{\rm 0,V}$ (Plummer) & $28\fm2 \pm 0\fm5$\\
 $\mu_{\rm 0,V}$ (Exponential) & $27\fm8 \pm 0\fm5$\\
V$_{\rm tot}$ & $13\fm9 \pm 0\fm5$\\
(m$-$M)$_0$ & $21\fm75 \pm 0\fm15$\\
 M$_{\rm tot,V}$ & $-7\fm9 \pm 0\fm5$ \enddata
\tablenotetext{a}{Surface brightnesses and integrated magnitudes are
corrected for the mean Galactic foreground reddenings, A$_{\rm V}$,
shown.}
\label{tab:struct}
\end{deluxetable}

\section{Data and Discovery}

The SDSS is an imaging and spectroscopic survey.  SDSS
imaging data are produced in five photometric bands, namely $u$, $g$,
$r$, $i$, and $z$~\citep{Fu96,Gu98,Ho01,Am06,Gu06}, 
and are automatically
processed through pipelines to measure photometric and astrometric
properties \citep{Lu99,St02,Sm02,Pi03,Iv04}. DR5 primarily covers $\sim 8000$ square degrees around the North Galactic Pole (NGP).

In the process of analyzing this area of the NGP~\citep{Be06b}, we visually identified a stellar overdensity in the
constellation Canes Venatici. The upper left panel of
Figure~\ref{fig:cvn_disc} shows a grayscale SDSS image of the sky
centered on the stellar overdensity; note that no obvious
object can be seen. However, a roughly elliptical overdensity of objects
classified by the SDSS pipeline as stars is readily visible in the
photometric data (upper middle and right panels). Plotting these stars
in a color-magnitude diagram (CMD) reveals a clear red giant branch
(RGB) and horizontal branch (lower panels). This CMD appears similar
to that seen in SDSS data for the Sextans dSph \citep[see, e.g.,
Fig. 1 of][]{Wi05}, although considerably more distant.
Based on its apparent morphology and on the presence of a distinct
stellar population, we conclude that this is most likely a hitherto
unknown dSph galaxy. As is customary, we name it after its
constellation, Canes Venatici (CVn).

\section{Physical Properties and Stellar Population}

The distance of a distinct, metal-poor stellar population can be
estimated from the $I$-band magnitude of the tip of the red giant
branch \citep[TRGB; e.g.,][]{moul83,Lee93}. We applied the transforms from
\citet{Sm02} to convert dereddened SDSS $g,r,i$ magnitudes to $V,I$
magnitudes. The cumulative $I$-band luminosity function for the RGB
shows a sharp rise starting at $I \sim 17.75$ (which corresponds to $i
\sim 18.23$). The horizontal branch fit in the lower right panel of
Figure~\ref{fig:cvn_disc} provides supporting evidence for the
reliability of our procedure.  Assuming a metallicity of [Fe/H]$ \sim
-2$ (see below) and a TRGB color of $(V - I)_{\rm TRGB} =1.4$, from
the calibration of \citet{daco90} we obtain the relation $(m-M)_0 =
I_{\rm TRGB} + 3.97$, yielding a distance modulus of $(m-M)_0 \sim
21.75 \pm 0.20$ ($\sim 220_{-16}^{+25}$ kpc). This is comparable to the most distant Milky Way dSph satellites previously known, Leo I and II \citep[250 and 215 kpc, respectively:][]{lee93a,lee95}.

The age and metallicity of the stars in CVn can be estimated by
comparison with the CMDs of single epoch, single metallicity
populations, e.g., Galactic globular clusters. In the lower right
panel of Figure \ref{fig:cvn_disc}, we show a field-star-subtracted
Hess diagrams (color-magnitude density plots) in $(i,g - i)$. Fiducial
ridgelines of three globular clusters -- NGC 2419 ([Fe/H] $\sim
-2.1$), M~3 ([Fe/H] $\sim -1.6$), and M~71 ([Fe/H] $\sim -0.7$) -- are
overplotted from left to right. The best match is to NGC~2419, and in
combination with the prominent horizontal branch, the similarity
indicates that CVn is dominated by an old, metal-poor ([Fe/H] $\sim
-2$) stellar population. The horizontal branch extends very
far to the blue, although it is predominantly red. 
Such red-dominated horizontal branches, hinting at a range of ages in the stellar population, are also seen in other metal-poor dSphs \citep[e.g.,][]{harb01}.

\begin{figure}[t]
\epsscale{0.9}
\plotone{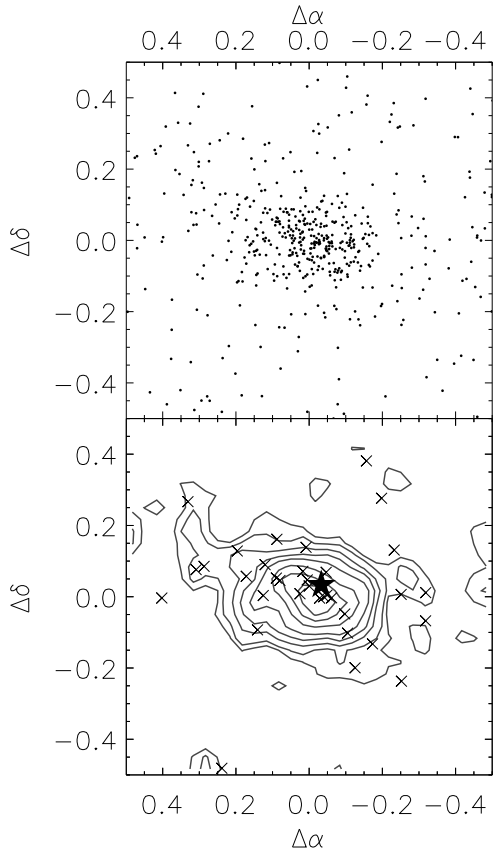}
\caption{Morphology of the CVn dwarf: {\it Upper:} The spatial
distribution of stars selected from CMD regions with the highest
contrast between CVn and field stellar populations. $\Delta \alpha$ and $\Delta\delta$ are relative offsets (arcdegrees) in right
ascension and declination. {\it Lower:} A
contour plot of the spatial distribution of these stars; candidate blue
horizontal branch stars (from the area blueward of the dashed line in
the lower right panel of Fig~\ref{fig:cvn_disc}) are overplotted with
crosses. The contours are 2, 3, 5, 7, 10, 15, 20, and 25$\sigma$ above the background level. The location of the carbon star, SDSS \starnum, is indicated
by the black star. \label{fig:cvn_morph}}
\vskip -0.20cm
\end{figure}


\begin{figure}[t]
\includegraphics[angle=-90,width=0.45\textwidth]{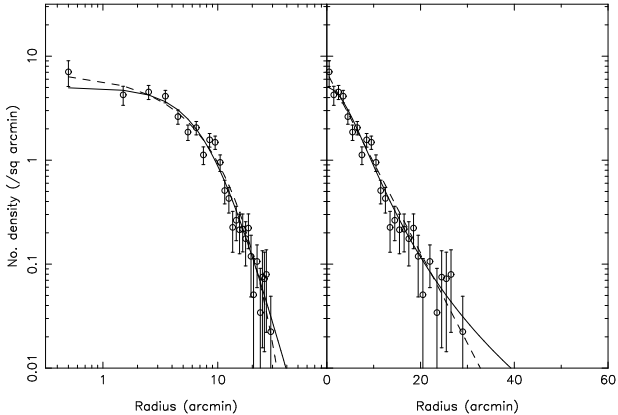}
\caption{Profile of CVn, showing the background-subtracted stellar density in elliptical
annuli as a function of mean radius. The left panel is logarithmic in
both axes, and the right panel is linear in radius. The overplotted
lines are fitted Plummer (solid) and exponential (dashed)
profiles.\label{fig:cvn_prof}}
\vskip -0.6cm
\end{figure}


\begin{figure}[t]
\epsscale{1.2}
\plotone{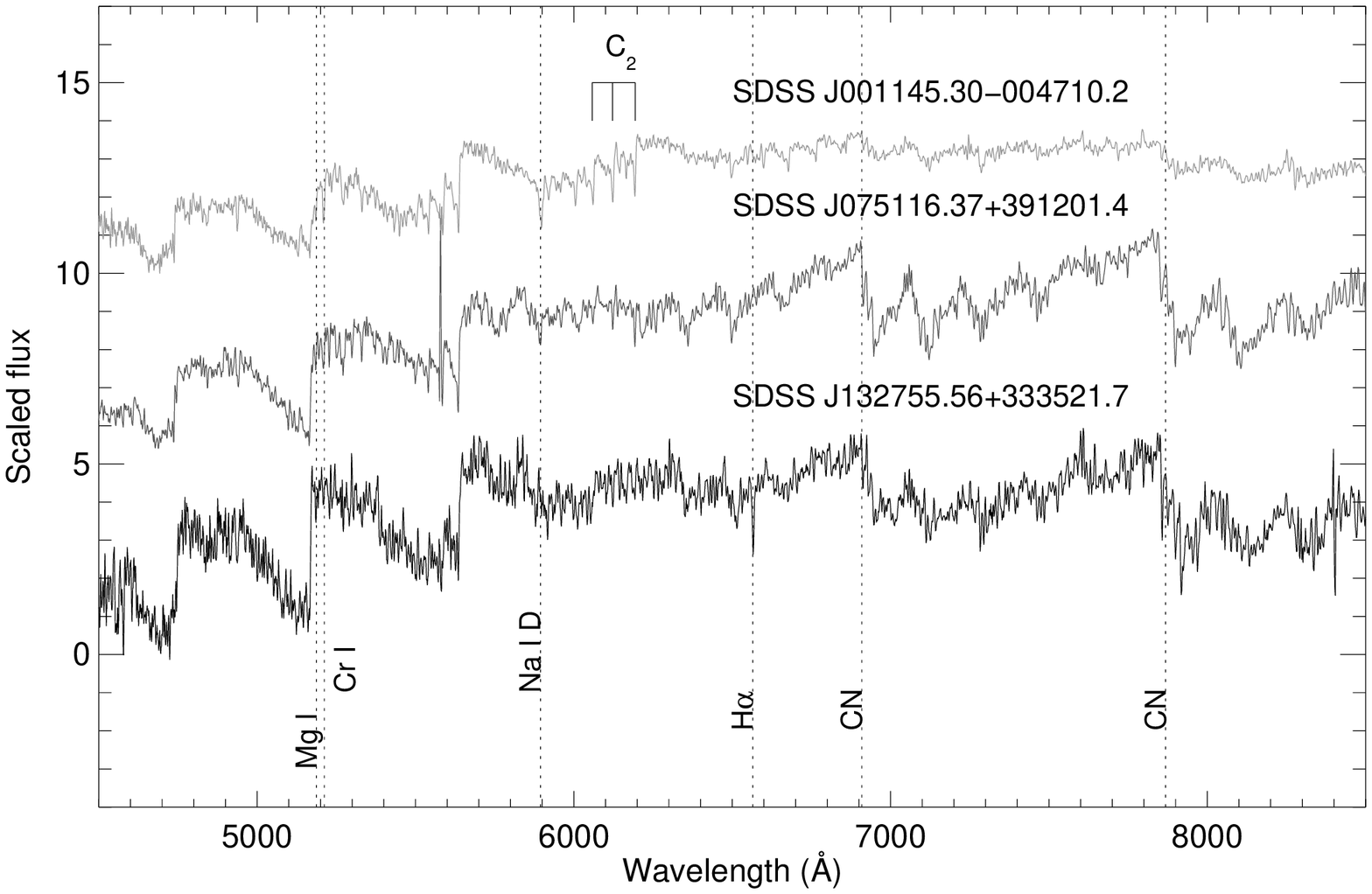}
\caption{Spectrum of SDSS\starnum \,({\it bottom}), a carbon star located near the
center of CVn (see lower left panel of Figure \ref{fig:cvn_disc} and
the lower panel of Figure~\ref{fig:cvn_morph}). Spectra of
a giant carbon star (SDSSJ075116.37+391201.4, {\it middle}) and a dwarf carbon star
(SDSSJ001145.30-004710.2, {\it top}) with similar temperatures \citep[T$_{\rm eff} \sim 4000$K;][]{down04} are shown for comparison. 
The CN bands are stronger in the reference giant than in the dwarf, while 
Na~I, Cr~I, and Mg~Ib lines and the C$_2$
bands around 6150 {\AA} are stronger in the dwarf. 
The strength of the former and weakness of the latter features in SDSS\starnum\, indicate that it is most likely a giant carbon star in CVn. 
\label{fig:spec} }
\vskip -0.7cm
\end{figure}

To investigate the spatial morphology, we select candidate members by applying
a mask built as follows. First, we find the class conditional
probabilities $P_{\rm CVn}$ and $P_{\rm bg}$ ~\citep[see
e.g.,][]{Be06a} of a star belonging to CVn and the background,
respectively. These are just the normalized Hess diagrams. The ratio
of the class conditional probabilities $x = P_{\rm CVn}/P_{\rm bg}$ is
converted to a probability of membership of CVn via $p = x / ( 1 +
x)$. A cut of $p > 0.8$ is used to select stars plotted in the top
panel of Fig.~\ref{fig:cvn_morph}. These objects are binned into $30 \times 30$
bins, each $0.033^\circ \times 0.033^\circ$, and smoothed with a Gaussian with FWHM of $0.067^\circ$ to yield the plot in the lower panel. The density contours, representing 2, 3, 5, 7, 10, 15, 20, and 25$\sigma$ above the background level, are
markedly irregular, particularly in the outer parts -- somewhat
reminiscent of the Ursa Minor dSph.  
The black crosses
are candidate blue horizontal branch stars (BHBs) selected with $g-i <
-0.1$. The distribution of BHBs is elongated in the same manner as the density
contours and traces the same underlying morphology.

To estimate the properties listed in Table~\ref{tab:struct}, we first
analyze the morphology shown in Figure~\ref{fig:cvn_morph} to derive
the centroid from the density-weighted first moment of the
distribution, and the average ellipticity and position angle using the
three density-weighted second moments \citep[e.g.,][]{St80}. The
radial profile shown in Figure~\ref{fig:cvn_prof} is derived by
computing the average density within elliptical annuli after first
subtracting a constant asymptotic background level (0.25
arcminute$^{-2}$) reached at large radii.  We then fit the radial
profile with standard Plummer and exponential laws
~\citep[Figure~\ref{fig:cvn_prof}, see also][]{Ir95}.  The
best-fitting position angle, ellipticity and half-light radii are
listed in Table~\ref{tab:struct}. At the distance of CVn, the
approximately $8\farcm5$ half-light radius corresponds to $\sim 550$ pc. The overall extent is $\sim 2$ kpc
along the major axis, making CVn one of the largest of the Milky Way
dSphs. Although Plummer and exponential laws provide reasonable fits
to the data, there are some discrepant datapoints in
Figure~\ref{fig:cvn_prof}~\citep[cf. Sextans' profile
in][]{Ir95}.

The overall luminosity is computed by masking the stellar locus of CVn
and computing the total flux within the mask and within the elliptical
half-light radius.  A similar mask, but covering a larger area to
minimize shot-noise, well outside the main body of CVn, is scaled by
relative area and used to compute the foreground contamination within
the half-light radius.  After correcting for this contamination, the
remaining flux is scaled to the total, assuming the fitted profiles
are a fair representation of the overall flux distribution. We also
apply a correction of $0.5$ magnitudes for unresolved/faint stars, based on the stellar luminosity functions of other low metallicity, low surface brightness dSphs. The resulting luminosity and central surface brightness estimates, M$_{\rm tot,V} \sim -7.9$ and $\mu_{\rm 0,V} \sim 28$ mag arcsec$^2$, are roughly consistent with the central surface brightness -- absolute magnitude relation noted for Local Group dSph satellites \citep[e.g.,][]{greb03}. However, CVn lies far from the central surface brightness -- host galaxy distance correlation found for other Local Group dSph satellites \citep[e.g., Fig. 7 of][]{mcco06}, suggesting that this correlation may be at least partially due to observational biases \citep[e.g.,][]{will04}.

A search of the SDSS DR5 spectroscopic database revealed the spectrum of a CH carbon star, SDSS \starnum,
in close proximity to the center of CVn (lower
panel of Figure \ref{fig:cvn_morph}). As shown in Figure \ref{fig:spec}, the star's spectral features indicate that it is most likely a giant, rather than a dwarf. There are no 2MASS
point sources within $15\arcsec$, and a comparison of the SDSS
coordinates with earlier-epoch APM data is consistent with zero proper
motion, both of which would argue in favor of its being a distant
object. Finally, the carbon star's dereddened
$r$-band magnitude is $19.18$; at the calculated distance modulus for
CVn, $\sim 21.75$, the star would have an absolute magnitude $M_r \sim
-2.5$, fairly typical of a CH carbon giant. Hence we conclude that
SDSS \starnum\, is a carbon star in CVn.
We measure a
heliocentric radial velocity of $+36 \pm 20$ km s$^{-1}$ for the carbon star from
cross-correlation with the spectrum of another CH carbon star of known
velocity~\citep[1249+0146, see][]{To89}. The carbon star's radial velocity in the
line of sight direction (after correction to the Galactic rest frame) is $\sim 80 \pm 20$ km s$^{-1}$.  
Given the distance of CVn, this is also approximately the radial velocity that
would be seen from the Galactic Center.  We plan to obtain follow-up
spectroscopy of RGB stars in CVn to supplement this first estimate of
its systemic velocity with a more detailed analysis of the new dwarf's
kinematics.

\section{Conclusions}

We have discovered a new dwarf spheroidal Milky Way satellite in the constellation Canes Venatici. It has an absolute
magnitude of $M_V \sim -7.9$ and a surface brightness of $\mu_{\rm
0,V} \sim 28$ mag arcsecond$^{-2}$. We find that CVn is dominated by
an old, metal-poor ([Fe/H] $\sim -2$) stellar population. At a
distance of $\sim 220$ kpc, CVn is one of the most remote of the known
dSph companions to the Milky Way.

It is important to carry out a well-defined census of Milky Way dwarf
galaxies to the faintest possible luminosities. Numerical simulations
in Cold Dark Matter cosmological models predict that the Milky Way's
halo should contain about 500 satellites comparable to the
dSphs~\citep{Mo99,Kl99}. This contrasts with the handful of known
Milky Way dSphs. The new dSph in Canes Venatici, taken with the recent
discovery of Ursa Major~\citep{Wi05}, emphasizes how incomplete our
current knowledge is. Further analysis of photometric data from large-area digital surveys such as SDSS will thus likely lead to the discovery of more dSphs around the Milky Way in the near future.

\vskip -0.8cm

\acknowledgments
DBZ, VB, MIW, DMB and MF acknowledge the financial support of the Particle Physics and Astronomy Research Council of the United Kingdom. 

Funding for the SDSS and SDSS-II has been provided by the Alfred P.
Sloan Foundation, the Participating Institutions, the National Science
Foundation, the U.S. Department of Energy, the National Aeronautics and
Space Administration, the Japanese Monbukagakusho, the Max Planck
Society, and the Higher Education Funding Council for England. The SDSS
Web Site is http://www.sdss.org/.                                              
                                 
The SDSS is managed by the Astrophysical Research Consortium for the
Participating Institutions. The Participating Institutions are the
American Museum of Natural History, Astrophysical Institute Potsdam,
University of Basel, Cambridge University, Case Western Reserve
University, University of Chicago, Drexel University, Fermilab, the
Institute for Advanced Study, the Japan Participation Group, Johns
Hopkins University, the Joint Institute for Nuclear Astrophysics, the
Kavli Institute for Particle Astrophysics and Cosmology, the Korean
Scientist Group, the Chinese Academy of Sciences (LAMOST), Los Alamos
National Laboratory, the Max-Planck-Institute for Astronomy (MPIA), the
Max-Planck-Institute for Astrophysics (MPA), New Mexico State
University, Ohio State University, University of Pittsburgh,
University of Portsmouth, Princeton University, the United States
Naval Observatory, and the University of Washington.




\end{document}